\documentclass[12pt,thmsa]{article}

\input tcilatex
\begin{document}

\title{\textbf{A PERTURBED KANTOWSKI-SACHS COSMOLOGICAL MODEL}}
\author{\ H. V. Fagundes \\
\ Instituto de F\'{\i}sica Te\'{o}rica, Universidade Estadual Paulista, \\
\ Rua Pamplona, 145, S\~{a}o Paulo, SP 01405-900, Brazil\\
\ e-mail: helio@ift.unesp.br \and M. A. M. Gonz\'{a}lez \\
\ Rua Homero Sales, 333, S\~{a}o Paulo, SP 05126-000, Brazil}
\maketitle

\begin{abstract}
A numerical integration is made, starting with a Kantowski-Sachs (KS) model
with an added initial perturbation at time one-tenth the age of the
universe, and evolving up to the present. According to a prediction by
Barrow (1989), the ratio between the fluctuations of the average matter
density and those of the metric tensor varies with the inverse squared
wavelength of the fluctuation. In this paper we take an extravantly large
value for this wavelength (20\% of the horizon's radius), to see that even
then that ratio remains greater than unity. So we may be reasonably
confident of models assuming density homogeneity, despite the fact that the
scale of observed structure in the universe has been increasing lately.
\end{abstract}

\section{Introduction}

This paper is a sequel to a study on the evolution of small perturbations on
cosmological models, either of the Friedmann-Lema\^{\i}tre-Robertson-Walker
(FLRW) family or closely related to it. The leading idea is to show that
small perturbations of the spacetime metric are compatible with not-so-small
perturbations of the average matter density in the universe - cf. Fagundes
\& Kwok (1991). Here we start with a Kantowski-Sachs (KS) model, which is
similar to the spherical FLRW model, and is easier than the latter to deal
with in a numerical study: while the space section for the latter is the
hypersphere $S^3$, for KS it is $S^2\times R$, where $S^2$\ is the ordinary
sphere and $R$\ is the real axis. Then we introduce deviations from the KS
metric at an initial time $t=0.1,$\ taking the age of the universe as $%
t_0=1. $\ The idea is to compare this KS model with perturbations which
evolve in time with the unperturbed KS universe.

In Fagundes \& Kwok (1991), the scale (wavelength) of the perturbation was
much smaller than the radius of observational horizon, and so, as expected
from Barrow's (1989) heuristic result, the fluctuations of the metric
components turned out to be much smaller than the fluctuation of the matter
density. Here we choose a perturbation with a very large wavelength (1/5 of
the horizon's radius), with the result that one of the metric components
still varies less than (about one-half as much as) the density. This result
reinforces our current confidence in models based on the assumption of
homogeneity of matter distribution, such as those of the FLRW family, since
the scale of observed inhomogeneity is believed to be smaller than that of
the perturbation assumed in this investigation.

\section{Einstein equations}

We choose units such that $c=G=t_0$\ $=1$, and assume a metric in the form 
\begin{equation}
ds^2=dt^2-a^2(t\text{,}r)dr^2-b^2(t\text{,}r)\sin ^2r\ d\varphi ^2-c^2(t%
\text{,}r)d\zeta ^2\text{ ,}  \label{metric}
\end{equation}
with $0\leq r\leq \pi ,0\leq \varphi \leq 2\pi ,-\infty \leq \zeta \leq
+\infty $, which is to be compared with the KS metric (Kantowski \& Sachs
1966) 
\begin{equation}
ds^2=dt^2-a_{KS}^2(t)(dr^2+\sin ^2r\ d\varphi ^2)-c_{KS}^2(t)d\zeta ^2\text{
.}  \label{KSmetric}
\end{equation}

For the metric given by Eq. \ref{metric} Einstein equations are (a dot means 
$\partial /\partial t$, a prime $\partial /\partial r$):

\begin{equation}
\frac{\dot{a}\dot{b}}{ab}+\frac{\dot{a}\dot{c}}{ac}+\frac{\dot{b}\dot{c}}{bc}%
+\frac 1{a^2}+\frac{a^{\prime }b^{\prime }}{a^3b}+\frac{a^{\prime }c^{\prime
}}{a^3c}-\frac{b^{\prime }c^{\prime }}{a^2bc}+(\frac{a^{\prime }}{a^3}-\frac{%
2b^{\prime }}{a^2b}-\frac{c^{\prime }}{a^2c})\cot r-\frac{b^{\prime \prime }%
}{a^2b}-\frac{c^{\prime \prime }}{a^2c}=8\pi \rho  \label{G00}
\end{equation}

\begin{equation}
\frac{\dot{a}b^{\prime }}{ab}-\frac{\dot{b}^{\prime }}b+\frac{\dot{a}%
c^{\prime }}{ac}-\frac{\dot{c}^{\prime }}c+(\frac{\dot{a}}a-\frac{\dot{b}}%
b)\cot r=-8\pi \rho v  \label{G01}
\end{equation}

\begin{equation}
\frac{\ddot{b}}b+\frac{\ddot{c}}c+\frac{\dot{b}\dot{c}}{bc}-\frac{b^{\prime
}c^{\prime }}{a^2bc}-\frac{c^{\prime }}{a^2c}\cot r=0  \label{G11}
\end{equation}

\begin{equation}
\frac{\ddot{a}}a+\frac{\ddot{c}}c+\frac{\dot{a}\dot{c}}{ac}-\frac{a^{\prime
}c^{\prime }}{a^3c}-\frac{c^{\prime \prime }}{a^2c}=0  \label{G22}
\end{equation}

\begin{equation}
\frac{\ddot{a}}a+\frac{\ddot{b}}b+\frac{\dot{a}\dot{b}}{ab}+\frac 1{a^2}+%
\frac{a^{\prime }b^{\prime }}{a^3b}+(\frac{a^{\prime }}{a^3}-\frac{%
2b^{\prime }}{a^2b})\cot r\ =0  \label{G33}
\end{equation}
where $\rho $\ $=\rho (t,x)$\ is the average (pressureless) matter density
and $v=v(t,x)$\ is the velocity field with respect to the Hubble flow,
assumed for simplicity to have only the $x$-component. Terms of the order of 
$v^2$\ are discarded.

As is well known (see, for example, Weinberg 1972, p. 163), Eqs. \ref{G00}
and \ref{G01} are constraints on the solutions of the second order
quasi-linear Eqs. \ref{G11}-\ref{G33}. To solve the latter we define new
variables $f=\dot{a}$, $g=\dot{b}$, and $h=\dot{c}$, in order to get a
system of six first-order (in $t)$\ differential equations for the dependent
variables $a,$\ $b,$\ $c,\ f,$\ $g,$\ and $h$. It is convenient also to
define $f_{KS}=\dot{a}_{KS},\ h_{KS}=\dot{c}_{KS}$.

We want our solution to as close as possible to FLRW's spherical model with
a dustlike matter distribution, so we take as our comparison KS solution one
with this property, namely that in Shikin's (1967) equations (16) and (20),
with $A=D/3=\bar{a}$, $B=E=0.$\ (See also Fagundes 1982.) The functions $%
a_{KS}$, $c_{KS}$\ are obtained in terms of the parameter $\eta \left(
t\right) $, which is the inverse of $t(\eta )=\bar{a}(\eta -\sin \eta )$, as
in the FLRW models. Here

\begin{eqnarray*}
a_{KS}(t)=\bar{a}[1-\cos \eta (t)], \\
c_{KS}(t)=3\bar{a}\{2-\eta (t)\cot [\eta (t)/2]\},
\end{eqnarray*}
and we choose $\bar{a}$\ $=[\pi /3-\sin (\pi /3)]^{-1}$, which corresponds
to $\eta (1)=\pi /3$,\ $\Omega _0^{FLRW}=4/3$, and $\rho
_{KS}(1)=[(4/3)a_{KS}(1)/c_{KS}(1)]\rho _0^{crit}\cong 1.1934557651\rho
_0^{crit}$, where $\rho _0^{crit}$\ is the usual critical density (present
density in Einstein-de Sitter's model).

\section{Numerical integration}

The system was then numerically integrated, in the interval $t=0.1$\ to $%
t=1.0$. (An attempt was made to start at $t=2\times 10^{-5},$\ the
recombination time; but the result was not satisfactory.) The initial
conditions are

\begin{eqnarray*}
a(0.1,r) &=&a_{KS}(0.1)[1+\alpha \cos (10r)] \\
b(0.1,r) &=&a_{KS}(0.1)[1+\alpha \cos (10r)] \\
c(0.1,r) &=&c_{KS}(0.1)[1+\beta \cos (10r)] \\
f(0.1,r) &=&f_{KS}(0.1)[1+\alpha \cos (10r)] \\
g(0.1,r) &=&f_{KS}(0.1)[1+\alpha \cos (10r)] \\
h(0.1,r) &=&h_{KS}(0.1)[1+\beta \cos (10r)]
\end{eqnarray*}
To get the desired effects we put $\alpha =\beta =0.02.$\ Note also the
large value of the initial perturbation comoving wavelength, $L=\pi /5$.
Despite this the perturbation remains sub-horizon sized: $\lambda (t)\approx
t^{2/3}L<3t$ for $t\geq 0.1$; so we need not worry about gauge ambiguities -
see, for example, Kolb \& Turner (1990).

The region of integration was divided into $1000$\ parts for the $r$\
variable, and $9000$\ parts for $t$, which means a spatial interval $\delta
r=10^{-3}\pi $\ and an integration step $\delta t=10^{-4}$. First and second
spatial derivatives were obtained by the usual rules, with a shift of $%
0.5\delta r$\ to soften the effect of the coordinate singularities at $%
r=0,\pi $: for example, for each $t=0.1+n\delta t,\,0\leq n\leq 9000$,
putting $r=(k+0.5)\delta r$, $0\leq k\leq 1000$, we defined $\ A[k]=$\ $%
a(t,r),$\ $B[k]=b(t,r)$, and so on; and took $a^{\prime }(t,r)$, $b^{\prime
\prime }(t,r)$\ respectively as

\[
AR[k]=(A[kPlus]-A[kMinus])/2\delta r, 
\]

\[
bRR[k]=(B[kPlus]-2B[k]+B[kMinus])/\delta r^2, 
\]
where

\[
(kPlus,kMinus)=\left\{ 
\begin{array}{c}
(1,0)\text{ for }k=0 \\ 
(k+1,k-1)\text{ for }0<k<1000 \\ 
(1000,999)\text{ for }k=1000
\end{array}
\right. . 
\]
After each increment $t\rightarrow t+\delta t$\ we make $a(t,r)\rightarrow
a(t,r)+f(t,r)\delta t$, and similarly for $b$, $c$, $f$, $g$, $h$. As for $%
\rho (t,r)$\ and $v(t,r)$\ they are given by the constraint equations.
Stability of the solution was checked by redoing the integration with $%
\delta \alpha ,\delta \beta =\pm 0.002$.

Calculations followed the simple rules, as given, for example, by Smith
(1978), and were programmed in the C language on an HP 750 workstation.

\section{Results}

The results at $t=1.0$\ are shown in Table 1, for uniformly spaced values of 
$r/\pi $. The entries are $a_{rel}=a(1,r)/a_{KS}(1)$, $%
b_{rel}=b(1,r)/a_{KS}(1)$, $c_{rel}=c(1,r)/c_{KS}(1)$, $v(1,r)$\ in km/sec,
and $\rho _{rel}=\rho (1,r)/\rho _{KS}(1)$. Except for $v(1,r),$\ they were
fitted to the following short Fourier series, where $x=r/\pi $:

\begin{eqnarray}
a_{rel}(x) &=&1.0008-0.1276\cos (10\pi x)+0.0051\cos (20\pi x),  \label{aRel}
\\
b_{rel}(x) &=&0.9932+0.0200\cos (10\pi x)+0.0069\cos (20\pi x),  \label{bRel}
\\
c_{rel}(x) &=&1.0008+0.0202\cos (10\pi x)-0.0008\cos (20\pi x),  \label{cRel}
\\
\rho _{rel}(x) &=&1.0236+0.2256\cos (10\pi x)-0.0041\cos (20\pi x),
\label{rhoRel}
\end{eqnarray}
These fits have the approximate periodicity of the initial perturbation, and
are symmetrical about $\rho =\pi /2$. Not counting the suspicious values
near $r=0$, $\pi $, note the fluctuations of about $13\%$\ for $a_{rel}$, $%
2\%$\ for $b_{rel}$\ and $c_{rel}$, and $25\%$\ for $\rho _{rel}$. Further
details are given in Gonz\'{a}lez (1994).

The fluctuation of $a_{rel}$\ is one-half that of $\rho _{rel}$,
qualitatively confirming Barrow's (1989) result, that $\delta \rho /\rho _0$%
\ $\sim (ct_0/L)^2\delta \phi /\phi _0$, where $\phi $\ is the Newtonian
potential (which corresponds to the metric tensor in general relativity),
and $L$\ is the scale of the perturbation. Here we have $L=(\pi /15)\times
3ct_0$, or about $20\%$\ of the horizon's radius; this makes Barrow's
estimate $\delta \rho /\rho _0$\ $\sim 2.53\times \delta \phi /\phi _0$. In
Fagundes \& Kwok (1991) we had an initial perturbation with wavelength $%
L=300h^{-1}$ Mpc, and a ratio of the order of $100$\ was obtained between $%
\delta \rho /\rho _0$\ and $\delta \phi /\phi _0$.

The fact that $\delta a_{rel}$\ is still smaller than $\delta \rho _{rel}$,
even with our exaggerated asumption for the size of $L$\ suggests that as
long as the scale of observed inhomogeneities is smaller than, say, $%
600h^{-1}$Mpc, models based on the homogeneity of matter distribuition are
reasonably correct as far as the metric tensor is concerned. However, this
is not an absolute conclusion. In Fagundes \& Mendon\c {c}a da Silveira
(1995), where initial conditions were contrived to illustrate another
problem, we got $\delta \rho /\rho _0\sim 2.5\times \delta \phi /\phi _0$\
for $L=120h^{-1}$\ Mpc.

\section{Acknowledgements}

One of us (M.A.M.G.) thanks Conselho Nacional de Desenvolvimento
Cient\'{\i}fico e Tecnol\'{o}gico (CNPq - Brazil) for a scholarship; H.V.F.
is grateful to CNPq for partial financial support.\vspace{1.0in}

\begin{center}
\ REFERENCES
\end{center}

Barrow, J.D. 1989, Quart. J.R.A.S., 30, 163

Fagundes, H.V. 1982, Lett. Math. Phys., 6, 417

Fagundes, H.V., \& Kwok, S.F. 1991, Ap.J., 368, 337

Fagundes, H.V., \& Mendon\c {c}a da Silveira, F.E. 1995, Brazil. J. Phys.,
25, 219

Gonz\'{a}lez, M.A.M. 1994, Universo de Kantowski-Sachs com Perturba\c
{c}\~{o}es,

\quad MS dissertation (S\~{a}o Paulo: IFT/UNESP); in Portuguese

Kantowski, R., \& Sachs, R.K. 1966, J. Math. Phys., 7, 443

Kolb, E.W., \& Turner, M.S. 1990, The Early Universe (Reading, MA:

\quad Addison-Wesley)

Shikin, I.S. 1967, Sov. Phys. Doklady, 11, 944

Smith, J.D. 1978, Numerical Solutions of Partial Differential Equations

\quad (2nd ed; Oxford: Clarendon Press)\qquad \qquad \qquad

Weinberg, S. 1972, Gravitation and Cosmology (New York: Wiley)

\newpage\ 

\vspace{1.0in}TABLE 1. Results of the integration for the functions defined
in the beginning \quad of Section 4.

\begin{center}
\begin{tabular}{ccrllc}
\hline\hline
$r/\pi $ & $a_{rel}$ & \multicolumn{1}{c}{$b_{rel}$} & \multicolumn{1}{c}{$%
c_{rel}$} & \multicolumn{1}{c}{$\rho _{rel}$} & $v(1,r)$ \\ \hline\hline
\multicolumn{1}{r}{0.0005} & \multicolumn{1}{r}{0.87197} & 0.87197 & 
\multicolumn{1}{r}{1.02292} & \multicolumn{1}{r}{1.56328} & 1077.29 \\ \hline
\multicolumn{1}{r}{0.0504} & \multicolumn{1}{r}{0.99773} & 0.90637 & 
\multicolumn{1}{r}{1.00227} & \multicolumn{1}{r}{1.15528} & 
\multicolumn{1}{r}{-69.32} \\ 
\multicolumn{1}{r}{0.1004} & \multicolumn{1}{r}{1.13339} & 0.98062 & 
\multicolumn{1}{r}{0.97988} & \multicolumn{1}{r}{0.79305} & 
\multicolumn{1}{r}{-20.79} \\ 
\multicolumn{1}{r}{0.1503} & \multicolumn{1}{r}{0.99464} & 1.03031 & 
\multicolumn{1}{r}{1.00157} & \multicolumn{1}{r}{0.96130} & 
\multicolumn{1}{r}{82.13} \\ 
\multicolumn{1}{r}{0.2003} & \multicolumn{1}{r}{0.87825} & 1.01964 & 
\multicolumn{1}{r}{1.02014} & \multicolumn{1}{r}{1.24580} & 
\multicolumn{1}{r}{5.93} \\ \hline
\multicolumn{1}{r}{0.2502} & \multicolumn{1}{r}{0.99636} & 0.98646 & 
\multicolumn{1}{r}{1.00129} & \multicolumn{1}{r}{1.02331} & 
\multicolumn{1}{r}{-69.62} \\ 
\multicolumn{1}{r}{0.3002} & \multicolumn{1}{r}{1.13342} & 0.98029 & 
\multicolumn{1}{r}{0.97970} & \multicolumn{1}{r}{0.79368} & 
\multicolumn{1}{r}{-7.77} \\ 
\multicolumn{1}{r}{0.3501} & \multicolumn{1}{r}{0.99506} & 1.00894 & 
\multicolumn{1}{r}{1.00148} & \multicolumn{1}{r}{0.99168} & 
\multicolumn{1}{r}{73.20} \\ 
\multicolumn{1}{r}{0.4001} & \multicolumn{1}{r}{0.87831} & 1.01973 & 
\multicolumn{1}{r}{1.02015} & \multicolumn{1}{r}{1.24548} & 
\multicolumn{1}{r}{1.47} \\ \hline
\multicolumn{1}{r}{0.4500} & \multicolumn{1}{r}{0.99572} & 0.99899 & 
\multicolumn{1}{r}{1.00138} & \multicolumn{1}{r}{1.00534} & 
\multicolumn{1}{r}{-70.74} \\ 
\multicolumn{1}{r}{0.5000} & \multicolumn{1}{r}{1.13343} & 0.98022 & 
\multicolumn{1}{r}{0.97969} & \multicolumn{1}{r}{0.79376} & 
\multicolumn{1}{r}{0.00} \\ 
\multicolumn{1}{r}{0.5500} & \multicolumn{1}{r}{0.99572} & 0.99899 & 
\multicolumn{1}{r}{1.00138} & \multicolumn{1}{r}{1.00534} & 
\multicolumn{1}{r}{70.74} \\ 
\multicolumn{1}{r}{0.5999} & \multicolumn{1}{r}{0.87831} & 1.01973 & 
\multicolumn{1}{r}{1.02015} & \multicolumn{1}{r}{1.24548} & 
\multicolumn{1}{r}{-1.47} \\ \hline
\multicolumn{1}{r}{0.6498} & \multicolumn{1}{r}{0.99506} & 1.00894 & 
\multicolumn{1}{r}{1.00148} & \multicolumn{1}{r}{0.99168} & 
\multicolumn{1}{r}{-73.20} \\ 
\multicolumn{1}{r}{0.6998} & \multicolumn{1}{r}{1.13342} & 0.98029 & 
\multicolumn{1}{r}{0.97970} & \multicolumn{1}{r}{0.79368} & 
\multicolumn{1}{r}{7.77} \\ 
\multicolumn{1}{r}{0.7498} & \multicolumn{1}{r}{0.99636} & 0.98646 & 
\multicolumn{1}{r}{1.00129} & \multicolumn{1}{r}{1.02331} & 
\multicolumn{1}{r}{69.62} \\ 
\multicolumn{1}{r}{0.7997} & \multicolumn{1}{r}{0.87825} & 1.01964 & 
\multicolumn{1}{r}{1.02014} & \multicolumn{1}{r}{1.24580} & 
\multicolumn{1}{r}{-5.93} \\ \hline
\multicolumn{1}{r}{0.8497} & \multicolumn{1}{r}{0.99464} & 1.03031 & 
\multicolumn{1}{r}{1.00157} & \multicolumn{1}{r}{0.96130} & 
\multicolumn{1}{r}{-82.13} \\ 
\multicolumn{1}{r}{0.8996} & \multicolumn{1}{r}{1.13339} & 0.98062 & 
\multicolumn{1}{r}{0.97988} & \multicolumn{1}{r}{0.79305} & 
\multicolumn{1}{r}{20.79} \\ 
\multicolumn{1}{r}{0.9496} & \multicolumn{1}{r}{0.99773} & 0.90637 & 
\multicolumn{1}{r}{1.00227} & \multicolumn{1}{r}{1.15528} & 
\multicolumn{1}{r}{69.32} \\ 
\multicolumn{1}{r}{0.9995} & \multicolumn{1}{r}{0.87197} & 0.87197 & 
\multicolumn{1}{r}{1.02292} & \multicolumn{1}{r}{1.56328} & -1077.45 \\ 
\hline\hline
\end{tabular}
\end{center}

\end{document}